\documentclass[a4paper,12pt]{article}
\usepackage{amssymb,amsmath,mathrsfs}
\usepackage[usenames,dvipsnames]{xcolor}
\usepackage{hyperref}
\usepackage{tensor}
\usepackage{graphicx}
\usepackage[left=1.2in,top=1in,right=1.2in,bottom=1in,headheight=0.8in,foot=0.5in]{geometry}
\usepackage[nodisplayskipstretch]{setspace} 
\usepackage[width=\textwidth]{caption}
\usepackage{pifont}% http://ctan.org/pkg/pifont
\usepackage[style=ext-authoryear-comp,maxcitenames=2,uniquename=false,backend=biber,uniquelist=true,articlein=false]{biblatex}
\DeclareNameAlias{sortname}{last-first}

\usepackage[indentafter]{titlesec}
\titleformat{name=\section}{}{\thetitle.}{0.8em}{\centering\scshape}
\titleformat{name=\subsection}[runin]{}{\thetitle.}{0.5em}{\bfseries}[.]
\titleformat{name=\subsubsection}[runin]{}{\thetitle.}{0.5em}{\itshape}[.]
\titleformat{name=\paragraph,numberless}[runin]{}{}{0em}{}[.]
\titlespacing{\paragraph}{0em}{0em}{0.5em}
\titleformat{name=\subparagraph,numberless}[runin]{}{}{0em}{}[.]
\titlespacing{\subparagraph}{0em}{0em}{0.5em}

\newcommand{\nocontentsline}[3]{}
\newcommand{\tocless}[2]{\bgroup\let\addcontentsline=\nocontentsline#1{#2}\egroup}
\hypersetup{
    breaklinks=true,
	colorlinks=true,         
	linkcolor=Black,          
	citecolor=MidnightBlue,
	urlcolor=MidnightBlue            
 }

\title{Cosmological Inflation and Meta-Empirical Theory Assessment\footnote{Forthcoming in \textit{Studies in History and Philosophy of Science}}}
\date{}

\author{William J.~Wolf\footnote{Faculty of Philosophy, University of Oxford, UK. Email: william.wolf@philosophy.ox.ac.uk}}

\makeatletter
\let\uppercasenonmath\@gobble

\begin{document}
\setstretch{1.0}
\maketitle

\begin{abstract}
\noindent I apply Dawid's Meta-Empirical Assessment (MEA) methodology to the theory of cosmological inflation. I argue that applying this methodology does not currently offer a compelling case for ascribing non-empirical confirmation to cosmological inflation. In particular, I argue that despite displaying strong instances of Unexpected Explanatory Coherence (UEA), it is premature to evaluate the theory on the basis of the No Alternatives Argument (NAA). More significantly though, I argue that the theory of cosmological inflation fails to sustain a convincing Meta-Inductive Argument (MIA) because the empirical evidence and theoretical successes that it seeks to draw meta-empirical support from do not warrant a meta-inductive inference to inflation.  I conclude by assessing how future developments could pave the way towards crafting a more compelling case for the non-empirical confirmation of cosmological inflation.
\end{abstract}
\tableofcontents
\setstretch{1.2}

\section{Introduction}

\noindent Developing methodological tools for the evaluation of scientific theories has long been a central theme in the philosophy of science. This particular area of the literature largely springs from Popper's famous analysis of the demarcation problem, where he proposed empirical ``falsification" as the criteria that drove scientific progress \parencite{Popper1935-POPTLO-7}. Other such programmes developed, including Kuhn's model of theory-choice, whereby one assesses theories on the basis of competing theoretical virtues \parencite{Kuhn1962-KUHTSO-3}, Lakatos' analysis of progressive and degenerating research programmes \parencite{Lakatos1970-POPFAT}, and Laudan's emphasis on problem solving and pursuit-worthiness within research traditions \parencite{Laudan1977-LAUPAI}. These programmes largely focused on direct empirical verification in their assessments of scientific theories. However, due in part to fundamental theoretical physics outpacing the technological capacities of experimental science, there has been a recent emphasis on non-empirical considerations in scientific evaluation.

\textit{Meta-Empirical Assessment} (MEA), a methodological programme initially developed by \textcite{Dawid2013-DAWSTA}, stands in contrast to the older classic methodologies that more closely focused on empirical assessment. MEA does not deny the primacy of empirical assessment and its ultimate necessity, but rather highlights the role of non-empirical considerations in theory assessment. To be more specific, empirical evidence ``consists of data of a kind that can be predicted by the theory assessed on its basis", whereas non-empirical evidence is the kind of evidence that ``supports a theory even though the theory does not predict the evidence" \parencite[p.36]{Dawid2013-DAWSTA}. For example, the standard model of particle physics cannot itself predict that no successful alternative to it will be found; however, the fact that no other compelling theoretical framework has emerged to challenge it, after nearly a century and vast cognitive resources being devoted to this problem, should increase our confidence in its efficacy. This is especially valuable in situations where we currently do not have adequate avenues to fully adjudicate the empirical merits of the theory or theories in question. The MEA programme's criteria for the non-empirical assessment of theories are primarily built on three arguments (to be discussed in more detail in the next section): the \textit{No Alternatives Argument} (NAA), the \textit{Unexpected Explanatory Coherence Argument} (UEA), and the \textit{Meta-Inductive Argument} (MIA). 

Cosmological inflation is a popular and promising theory that has emerged as an extension to the standard hot big bang model of cosmology. This theory proposes that the universe underwent a period of dramatic expansion very early on in its history. While its advocates highlight inflation's ability to cleanly and efficaciously account for long-standing puzzles in cosmology, as well as tout some empirically successful predictions, direct empirical confirmation of the theory has proven to be somewhat elusive despite expectations to the contrary from some members of the physics community. Inflation is thus an ideal candidate for Dawid's MEA programme. 

In this paper, I will evaluate the meta-empirical assessment prospects for the theory of cosmological inflation. I will argue that the theory has the potential to sustain an impressive UEA argument, but that it is somewhat premature to firmly assess it on the basis of an NAA argument as assessing the viability of alternatives has yet to fully play out within the physics community; which for the most part largely concurs with \textcite{DawidManuscript-DAWIIC} (see also \textcite{McCoy2021-MCCMSF-2}). However, I will disagree with Dawid and McCoy's assessment of the MIA argument as applied to inflation. In particular, I will closely examine arguably successful instances of MIA, including prominent examples from particle physics and cosmology, and demonstrate that inflation lacks the ingredients that makes these other MIA applications so compelling. In short, MIA-type reasoning for an unconfirmed scientific theory is most compelling when such a theory is strongly implicated by consistency arguments coming from the empirical evidence and the other well-confirmed scientific theories that it is drawing meta-empirical support from. Inflation, as I shall argue, does not quite fall into this category. While the theory has genuinely impressive achievements, its relationship to the theories that it seeks to draw MIA support from (the standard models of particle physics and cosmology) is driven primarily by explanatory considerations rather than by consistency arguments. This is significant because non-empirical confirmation of the type that the MEA programme envisions requires support from at least two (and preferably all three) arguments in order to get inference to the best explanation (IBE) or Bayesian reasoning off the ground so as to place non-trivial limitations on scientific underdetermination. Thus, I take the difficulties inflation has in multiple MEA dimensions to indicate that there is not (yet) a compelling non-empirical case for the theory.

This paper proceeds as follows. In section \S\ref{s2}, I recall the basics of the MEA programme and its original application to string theory. In section \S\ref{s3}, I draw particular attention to the MIA argument by exploring arguably successful examples of this reasoning playing out in both particle physics and cosmology. In section \S\ref{s4}, I apply MEA to inflationary cosmology, arguing that while it is premature to make an NAA-style argument for inflation due to currently viable alternatives, inflation does have an impressive UEA case. Furthermore, I argue that inflation has serious deficiencies with an MIA-tpye argument that is applied in its favor because in this context inflation's succeeds primarily on its explanatory benefits, rather than through consistency-driven, empirically-based inferences. In \S\ref{s6}, I argue that while applying Dawid's programme to inflationary cosmology does not offer compelling meta-empirical confirmation of inflation, future developments could conceivably make the case for meta-empirical confirmation significantly stronger. I conclude in \S\ref{s7}.

\section{Meta-Empirical Assessment} \label{s2}

\noindent MEA is comprised of three main arguments that work together to offer non-empirical confirmation to the theory under consideration. These are the \textit{No Alternatives Argument} (NAA), the \textit{Unexpected Explanatory Coherence Argument} (UEA), and the \textit{Meta-Inductive Argument} (MIA) \parencite{Dawid2013-DAWSTA}. In this section, I will review these arguments and how they collectively place limits on scientific udertermination, as well as examine how they are applied to string theory, the scientific theory that has been most often discussed in the context of this methodological programme.

\subsection{Meta-Empirical Assessment Basics}

\begin{enumerate}
    \item \textit{No Alternatives Argument:} This argument holds that, when assessing a scientific theory, it is instructive to consider the number of potential alternatives that can offer satisfying accounts of the same phenomena, while also remaining coherent and consistent with other theories and general background knowledge. That is, one conjectures ``a connection between the spectrum of theories the scientists came up with and the spectrum of all possible scientific theories that fit the available data [...] if a viable scientific theory exists and only very few scientific theories can be built in agreement with the available data, the chances are good that the theory actually developed by scientists is viable" \parencite[p.~51]{Dawid2013-DAWSTA}. If, after a long, exhaustive search, no other alternatives have emerged, this can indicate that there simply may not be viable alternatives that account for the phenomena in an equally satisfying manner.

    The obvious vulnerability with this line of reasoning is the prospect of unconceived alternatives. The human capacity for imagination, reasoning, and technical skill is certainly not infinite, leaving open the possibility that we are simply overlooking viable or even superior alternatives. In order for NAA to be a convincing argument, it needs to be supplemented with additional arguments in favor of the particular theory in question.
    \item \textit{Unexpected Explanatory Coherence Argument:} This argument points to instances where the theory in question provides additional explanatory power or enhances overall coherence with other theories and background knowledge, over and above what the theory was introduced to account for. This is a powerful argument in a theory's favor because it ``mirrors the canonical reasoning for a theory's viability based on novel empirical confirmation" \parencite[p.~52]{Dawid2013-DAWSTA} in that it provides novel explanatory power and coherence at the conceptual level and integrates the theory in question more fully with other phenomena and theories in our scientific background knowledge. An example of this would be the introduction of gauge symmetry into the standard model of particle physics. Gauge symmetries were introduced initially to solve the renormalization problem, but then also happened to provide a framework from which the entire spectrum of elementary particles could be explained and derived from purely theoretical arguments \parencite[p.~81]{Dawid2013-DAWSTA}.

    \item \textit{Meta-Inductive Argument:} This argument most closely resembles traditional models of empirical assessment. Here, one uses the empirically confirmed successes of other theories, models, or principles within the more general research programme or the empirically confirmed successes of particular components of the theory itself to infer that a theory is on the right track. There is a long history within the physical sciences of applying specific physical principles, problem solving techniques, and patterns of reasoning, and the empirical successes of these strategies speak to their viability even if a direct empirical test is not immediately available. ``The empirical observations that provide the basis for MIA thereby increase the trust in so far empirically unconfirmed scientific theories which are supported by the given strategies" \parencite[p.~53]{Dawid2013-DAWSTA}. This amounts to making a ``meta-inductive inference that regular predictive success in a research field justifies the assumption that future predictions of a similar kind will be correct as well. To be applicable the inference must rely on a reasonable understanding as to what can count as a prediction of a similar kind" \parencite[p.~55]{Dawid2013-DAWSTA}. 
    %This inference occurs at the meta-level as the actual, consistent predictive successes generated by particular theories or physical principles are not something that the theories themselves can predict, but nonetheless increase our trust in unconfirmed theories that also make similar applications of them. 

\end{enumerate}

It is clear that these arguments are all mutually re-enforcing and become far more compelling when they are stacked together. Yet, it is important to stress that none of them is significant in complete isolation from the others. Indeed, history is replete with examples of theories that seemed either to be the only game in town, or offered impressive explanatory power, or were similar to previously successful strategies, but later turned out to be wrong. Crucially, ``meta-empirical assessment needs to be based on at least two if not all three arguments in conjunction" in order to generate significant meta-empirical confirmation \parencite[p.~344]{DeBaerdemaeker:2022jkz}. 

Individually, we can understand all of these arguments as placing some constraints on scientific underdetermination, or the landscape of possible theories that can adequately account for our observations. NAA explicitly limits scientific underdetermination by arguing that there is no viable alternative that can adequately account for the same phenomena. UEA and MIA can as well, but in a more restricted sense. For example, instances of UEA can indicate limits on scientific underdetermination within the theory's regime of applicability (i.e.\ the landscape of possible theories within that regime is limited to those that retain certain explanatory features), but does not necessarily rule out a more fundamental theory from retaining these merits. Similarly, MIA puts limits on scientific underdetermination because the number of theories that are potentially compatible with successfully established empirical or theoretical knowledge will necessarily be restricted in some ways as every further instance of confirmation rules out the potential theories that are not compatible with each subsequent empirical confirmation. Taking instances where these arguments are all present forces the scientist to question the plausibility that a particular theory, one that seems to have no viable alternatives that account for the phenomena in a satisfactory manner, and that offers remarkable unexpected explanations, and that benefits from empirical support at the meta-level, could actually turn out to be mistaken. We can intuitively understand that this will begin to strain credulity at some point after enough of this kind of ``non-empirical" evidence has accumulated. 

When at least two (and better yet, all three) of these arguments are present, Dawid argues that they collectively provide compelling instances of non-empirical theory confirmation. He argues that we can understand such confirmation as either \textit{inference to the best explanation} (IBE) or \textit{Bayesian confirmation}. IBE infers the viability of statements based on the criteria that they offer the best explanation for the observations in question \parencite{Lipton2007-LIPITT-7, Bird2007-LIPITT-6}. This applies to non-empirical theory assessment because placing significant limitations on scientific underdetermination through NAA, UEA, and MIA dramatically increases the likelihood of the theory in question providing the best explanation for the phenomena \parencite[p.~65]{Dawid2013-DAWSTA}. Typical Bayesian reasoning holds that empirical data that supports the theory in question raises the probability of the theory's viability. This process proceeds iteratively and is continually updated as more evidence comes in. As has been argued by \textcite{Dawid2015-DAWTNA}, the core components of Bayesian reasoning function even if the evidence is not of the empirical kind. We can thus understand the limitations that NAA, UEA, and MIA place of scientific underdetermination to constitute the kind of evidence that causes us to update our probabilities regarding a theory's viability.\footnote{\textcite{Menon2019-MENOTV} has argued that this Bayesian reasoning does not apply in the case of the NAA due to worries that obtaining significant, or ``non-negligible confirmation", requires an implausible fine-tuning of the theory's priors. \textcite{pittphilsci18818} has responded by arguing that the ``priors needed for making
a no-alternatives argument significant are in line with what can be plausibly assumed in a successful research field".}

\subsection{MEA and String Theory}

Dawid initially developed MEA to account for both the strong degree of confidence that string theory has within the particle physics community and the comparatively cautious (to put it generously) attitude that other communities within the physical sciences have towards it. His analysis argues that the particle physics community had long been successfully using this kind of reasoning in developing the standard model of particle physics, which explains their relative confidence in the ultimate viability of string theory. As much of the literature surrounding this programme has been developed in the context of this string theory application, it will be helpful to briefly recall how these arguments are applied to argue for the viability of string theory as a way of seeing the programme in practice. 

As with any discussion of string theory, perspectives are highly divergent. Dawid, as well as the string community more generally, maintain that string theory convincingly sustains a uniquely strong NAA argument. It aspires to a universal description of all known interactions in terms of the contemporary particle physics research programme. As is well-known, gravity is a non-renormalizable interaction, but upon dropping the idea of point particles and positing the extendness of elementary particles (strings), one can use the traditional methods of particle physics to universally describe all known interactions down to the Planck scale; furthermore, there seems to be no other viable way to accomplish this. Rovelli notes though, the strength of this NAA argument and the idea that string theory is the only game in town is very dependent on the set of assumptions a theorist is working under. As a researcher in such an alternative programme, he points out that ``an alternative to string theory is loop quantum gravity, considered the “only game in town” by those who embrace it, under their set of assumptions" \parencite[p.~2]{Rovelli:2016cpd}. What are the assumptions that give string theory the most plausible NAA claim? Crucially, there are good reasons to believe that moving from point particles to strings is the only way of extending the enormously successful quantum field theory principles and techniques to a theory that unifies and encompasses all four known interactions, while respecting fundamental principles such as causality and unitarity \parencite{Polchinski:1998rq, Polchinski:1998rr}. That is, if one assumes that the universe and all of its interactions down to the most fundamental levels are correctly described by the gauge-symmetric, quantum theoretic principles used in construction of the standard model of particle physics, string theory does seem have an interesting NAA case. However, this assumes that none of the above principles need adjusting and that there are not further, as of yet unknown fundamental principles that become relevant at these scales.  

Moving on to UEA, string theory is noted for producing a significant number of instances of unforeseen explanations and generating coherence with other areas of physics. For example, string theory not only makes gravity renormalizable, but also implies that the graviton, along with other fundamental particles, naturally emerges in a unified framework as different oscillation modes of the string \parencite[p.~33]{Dawid2013-DAWSTA}. As another example, black hole entropy can be understood within the string framework by counting the microstates in the string theoretic description of certain black holes, thus generating coherence with thermodynamic principles that have always been puzzling when applied to black holes and gravitational phenomena \parencite{Strominger:1996sh}. 

The MIA argument for string theory is somewhat similar to the NAA argument because it follows from trusting the gauge and quantum field theory principles that are weaved into the standard model of particle physics. String theory emerged out of applying principles from quantum field theory to the study of fundamental interactions; these quantum field theory techniques and principles have seen spectacular success in the empirical confirmations of all the major components of the standard model. The empirical success of this programme in its totality provides non-empirical evidence for the viability of extending this programme to further interactions and higher energies, even if direct empirical tests of these extensions are not immediately available. However, it should also be pointed out that string theory has run into some difficulties in recent years. While not fatal blows, the failure to observe supersymmetric particles and the theory's (seemingly) generic prediction of a \textit{negative} cosmological constant have put pressure on the theory \parencite{Rovelli:2016cpd}.\footnote{The issue surrounding whether or not string theory allows for the construction of metastable de Sitter vacua, in contrast to the anti de Sitter vacua that lead to a negative cosmological constant, is contentiously debated within the string theory community to this day. See \textcite{Danielsson:2018ztv, Cicoli:2018kdo} for reviews from both sides of this debate.}

Furthermore, that the standard model can provide meta-inductive support to string theory at all has been disputed by \textcite{Chall2018-CHADFD-2}, who argues that MIA should not apply to successor theories because the empirical confirmation of the predecessor has already been accounted for as any successor theory \textit{must} retain the successes of its predecessor. It should not double count as meta-empirical support. Thus, in the case of string theory, the success of the standard model would not offer MIA support to a successor theory such as string theory. In a response, \textcite{pittphilsci18818} points out that meta-level evidence is qualitatively different than empirical results that are predicted by a predecessor theory because this meta-level evidence represents contingents facts that about the success of the research programme and underlying principles, facts that cannot be predicted by the either predecessor or successor theory.

Unsurprisingly, there is not a consensus regarding the degree to which string theory succeeds in its non-empirical (and empirical) merits. However, this discussion illustrates how these meta-empirical arguments function in the case of non-empirical theory assessment. As the MIA will feature particularly in this paper, we now turn to examine it a little more closely.

\section{Exploring the Meta-Inductive Argument} \label{s3}

\noindent Recall that MIA is an \textit{empirical} argument. That is, there must be a non-trivial, consequential connection between the hypothesis or theory that one would like to infer support for on the meta-level and the actual empirical evidence that one is citing in this inference. Crucially, ``the inference must rely on a reasonable understanding as to what can count as a prediction of a similar kind" \parencite[p.~55]{Dawid2013-DAWSTA}. As the MIA case for string theory extrapolates from a predecessor theory to a successor theory at a completely different scale of fundamentality, this is not the most useful example of this kind of inference for the theory we are going to assess because inflation is not a more fundamental theory, but rather an extension of the current standard model of cosmology. In this section, we will explore examples of this inference that are more readily comparable to cosmological inflation, including the Higgs mechanism and prior instances of MIA within the field of cosmology itself. As we shall see, these examples cash out this inference between empirical evidence and meta-level support for an unconfirmed theory in particularly compelling fashion. More specifically, the empirical evidence is of such a nature that the unconfirmed hypothesis we infer meta-level support for is naturally implicated by a consistent application of the confirmed parts of the theory.

\subsection{MIA Example 1: The Higgs Mechanism}

The so-called Higgs mechanism was independently proposed by a number of researchers \parencite{Higgs:1964pj, Englert:1964et, PhysRevLett.13.585} and performs a critical role in the standard model of particle physics as it does nothing less than account for the observed mass spectrum of the elementary particles. 

In brief, the concept of gauge symmetry is fundamental to the standard model of particle physics as these symmetries are crucial for constructing renormalizable quantum field theories that allow us to make empirical predictions concerning particle interactions. Additionally, it provides a framework from which the spectrum of existing particles can be derived from the group representations attached to these gauge symmetries. Gauge symmetries (on a standard physics interpretation of the concept) are local symmetries, or symmetries that can vary from spacetime point to spacetime point such as the U(1) symmetry of electromagnetism.\footnote{This is in contrast to global symmetries, such as a rigid Galilean transformation, which act identically on each spacetime point. See \textcite{MurgueitioRamirezForthcoming-MURAGS-3, Wolf:2021ydy, Wallace2014-GREECO} for some discussions in the philosophical literature concerning the interpretation and empirical significance of local and global symmetries.} When constructing the Lagrangian of the standard model, it turns out that introducing typical mass terms for elementary particles like electrons or W bosons spoils these gauge symmetries. This is incredibly inconvenient when all of these particles turn out to, in fact, possess mass. 

This problem was brilliantly solved with the addition of the Higgs field, a spin-zero scalar field, that couples to these particles. It preserves the important gauge symmetries present in the theory because this addition does not directly involve adding a mass term for any of the other fields in the theory's Lagrangian. However, it is energetically favorable for the potential of the Higgs field to rest at its minimum. When the field is at its minimum, it acquires a vacuum expectation value that, when realized in the theory, is an instance of spontaneous symmetry breaking. This vacuum expectation value of the Higgs field, through its coupling to the other fields in the theory, confers mass to the elementary particles. In particular, the way this mechanism is implemented preserves a massless photon (as expected), while providing a realistic mass spectrum to fermions and the rest of the bosons. 

As this theory percolated within the particle physics community, physicists developed an extraordinary degree of confidence, or even certainty, in the viability of the Higgs mechanism, to the point that they invested billions of dollars and decades of effort in the Large Hadron Collider (LHC) to discovery the Higgs boson and explore its properties. Failure to find the Higgs boson would have been a full-blown catastrophe for the particle physics community. 

Dawid argues that their confidence in the eventual confirmation of the Higgs mechanism comes about precisely through the types of non-empirical arguments used in his MEA methodology \parencite[p.~113]{Dawid2013-DAWSTA}. According to Dawid, the reasoning proceeds along the following lines. (i) Physicists needed quantum field theory to describe relativistic phenomena on the atomic and sub-atomic scales. (ii) Calculating particle interactions required renormalizable theories. (iii) Renormalizable theories necessarily possess a gauge symmetric structure. (iv) Gauge symmetric quantum field theories together with the existence of massive particles, require spontaneous symmetry breaking via the Higgs mechanism. Every one of these preceding steps (i-iii) involved spectacular empirical confirmations and the particles involved were known to possess mass. In the absence of any viable alternatives, it was clearly epistemically warranted to extrapolate from these empirical successes of the standard model and have a high degree of confidence in the viability of the Higgs mechanism even before obtaining direct empirical verification of the Higgs itself. As we all know, the Higgs was recently confirmed and this event signaled the triumph of the standard model of particle physics \parencite{ATLAS:2012yve}\footnote{See \textcite{Dawid2015-DAWHDA} for a discussion on some issues concerning the theoretical reasoning at play in interpreting the data that was used in this confirmation.}. This serves as the paradigmatic example of Dawid's MEA programme at its best and the MIA argument in particular.

What is the exact nature of this inference and what relationship do the prior empirical successes of the standard model have towards the unconfirmed Higgs theory? Consider \textcite[p.~112-113, my emphasis]{Dawid2013-DAWSTA}:
\begin{quote}
    ``\textit{The standard model just could not account for the occurrence of massive objects in the observed world if the Higgs sector was simply left out. Thus, the Higgs mechanism was an essential part of the standard model} and did not constitute an independent new theory in its own right...Nevertheless, it was based on a separable set of theoretical posits and its predictions could be distinguished from those based on other segments or principles of the standard model: it predicted at least one new particle and a certain structure of that particle's interactions with itself and with matter. The Higgs sector therefore constituted a separable ``module" of the standard model, a kind of sub-theory whose viability could be discussed separately from the other parts of the standard model." 
\end{quote}
The part of the quote that I have emphasized identifies that the introduction of the Higgs mechanism was a matter of \textit{empirical adequacy} for the standard model of particle physics. In other words, there was an \textit{inconsistency} between the empirical success of the gauge symmetric quantum field theories of the standard model and the empirically observed mass spectrum of elementary particles that could only be reconcilled with the Higgs mechanism. The standard model without the Higgs mechanism is not empirically adequate because it cannot account for the mass spectrum observed in fermions and bosons. Without the Higgs mechanism, the standard model describes these particles as massless, which plainly contradicts reality. It is not an exaggeration to say that the standard model would need to be abandoned without this modification. The mere fact that many of the most important predictions derived from the standard model can be aligned with the empirical data coming from particle physics experiments at all is inextricably dependent on the role that the Higgs field plays in the structure of the standard model. 

To be even more explicit regarding how this MIA inference works, the standard model of particle physics with the Higgs mechanism $SMP_H$ (standard model with the Higgs boson) is constructed using all of the insight, principles, and strategies used in gauge symmetric quantum field theory. The empirical data $O_{BH}$ (observations before the discovery of the Higgs boson) represents all the rich phenomenology explored in our particle accelerators before the Higgs boson was directly detected. In a traditional empirical methodology, $SMP_H$ is understood to predict $O_{BH}$ and the observations $O_{BH}$ are taken to offer confirmation to the particular modules of $SMP_H$ that are directly responsible for them (for example, discovering a particle based on observing its tracks and decay products in a scattering experiment would be seen as evidence for that particular component of the standard model). MIA holds that empirical observations $O_{BH}$ are also understood to offer non-empirical confirmation to the general research programme as a whole because they validate the principles and components used in the whole construction, including the modules of $SMP_H$ that did not yet have direct empirical confirmation $O_H$ until the 2012 discovery. What is the connection between $O_{BH}$ and $H$ that justifies such an inference? This particular MIA inference between the empirical evidence $O_{BH}$ and the unconfirmed module $H$ is underwritten by the following: the fact that the predictions of $SMP_H$ and the observations $O_{BH}$ coincide is necessarily contingent on $H$. In other words, $SMP$ does \textit{not} predict a significant portion of $O_{BH}$ without the addition of $H$. Drawing on $O_{BH}$ to make a meta-inductive inference for $H$ is thereby justified because $H$ is required for the empirical adequacy of the standard model that explains and predicts  $O_{BH}$. There is no consistent, empirically adequate standard model of particle physics without the Higgs mechanism. Even if direct observations of the Higgs mechanism took decades to produce, a significant portion of the empirical observations made in testing the standard model up to that point necessarily implicated the Higgs mechanism because it was the only way that those particular facets of the standard model could consistently be reconciled with the existence of the relevant mass spectrum of observed particles.

\subsection{MIA Example 2: Applying General Relativity to Cosmology}

Jim Peebles, a recent Nobel laureate and one of the preeminent figures in modern cosmology, has identified that non-empirical assessment played a major role in the history of cosmology both before and during his time in the field. Primarily, these meta-empirical inferences have involved trusting General Relativity (GR) over distance scales and in scenarios that were many orders of magnitude separated from the regimes that it was well-established in \parencite[Ch.~1]{Peebles+2020}. Essentially, cosmology has taken a theory that was tested over solar system and terrestrial-type scales and extrapolated it more than 15 orders of magnitutde to describe the dynamics and evolution of the known universe. For the purpose of exploring how such inferences have worked in cosmology, I have identified two of many potential instances where non-empirical, meta-level support has been inferred from an empirically well-confirmed theory: the realization that the universe evolves  dynamically and the existence of gravitational waves.\footnote{It should be noted here that MIA was initially intended to apply to inferences involving the non-empirically confirmed theories and \textit{other} empirically confirmed theories. However, I am grouping these novel applications of GR under MIA because the operative concept that is most important here is the \textit{inductive risk} taken in extrapolating these theories across so many orders of magnitude, where the idea is that this inductive risk mirrors the inductive risk present when making an inference from the empirical successes of another theory or model in the research programme that utilizes similar strategies or concepts. It is this inductive risk that Peebles is referring to when he discusses the meta-empirical considerations used in cosmological research. Furthermore, if we think of GR as a research programme rather than just a theory, these novel applications to such different phenomena over vast orders of magnitudes are plausibly imagined as sub-modules of the overall relativistic research programme. Once confirmed though, these instances would count as examples of novel confirmation and could thus be categorized under the UEA argument. Before confirmation though, I would argue that they bear most resemblance to MIA due to the inductive risk present in this kind of reasoning.}

\subsubsection{The Dynamical Universe}
Modern cosmology developed out of the theory of General Relativity, where \textcite{1917SPAW.......142E} proposed the first relativistic model of the universe in 1917.\footnote{See \textcite{ORaifeartaigh:2017uct} for a historical review.} This was his static model of the universe, (in)famous for featuring his supposed ``biggest blunder", the cosmological constant. Yet, dynamically evolving models of the universe were soon independently suggested by \textcite{1922ZPhy...10..377F, 1924ZPhy...21..326F} and \textcite{1927ASSB...47...49L}. In deciding between these two hypotheses, that of a static universe (the received view) and that of a dynamic universe, we can see evidence of MIA-type reasoning. The key realization, arguably obvious in Lemaitre's work, but first explicitly stated by \textcite{Eddington:1930zz}, was that Einstein's static universe is perturbatively unstable. This insight immediately renders the static universe phenomenologically unviable description of the universe. 

The reasoning proceeds as follows. GR had received impressive confirmation in a number of the classical tests of the theory, including the explanation of Mercury's perihelion and the prediction of the deflection of light rays observed in the 1919 total solar eclipse\footnote{This of course sets aside potential underdetermination issues in GR stemming from the existence of the geometric trinity: empirically equivalent gravitational theories formulated in terms of the different geometric concepts of curvature, torsion, or non-metricity. See e.g.\ \textcite{Wolf:2023xrv, Wolf:2023rad, Capozziello:2022zzh, BeltranJimenez:2019esp} for further discussions.}. This of course instills confidence in the general relativistic research programme and in its applicability to other, as of then untested, regimes. In applying GR to cosmological solutions, it becomes clear that static solutions are unstable and thus cannot represent an empirically viable description of the universe. Thus, we are confronted with two ideas: (i) GR's empirical confirmations within the solar system justify our confidence in extending the theory towards applications in other regimes, despite the fact that it had then not been tested at all within other regimes and (ii) a consistent application of GR to cosmological descriptions of the universe necessarily implies that the universe must evolve dynamically.

\textcite{Einstein1931} eventually proposed his own dynamical model of the universe, but also noted that the demonstration that his previous static model was unstable was sufficient grounds to reject it entirely independent of any other considerations; saying that ``on these grounds alone, I am no longer inclined to ascribe a physical meaning to my former solution, quite apart from Hubble’s observations" (translation from \textcite{ORaifeartaigh:2013khl}). The reference to the famous \textcite{hubble1929} results concerning the apparent recession of spiral nebulae might make it seem as if Einstein is using a more tradition kind of empirical reasoning here. But, as \textcite[p.~677]{Eddington:1930zz} pointed out, ``the proof of the instability of Einstein’s model greatly strengthens our grounds for interpreting the recession of the spiral nebulae as an indication of world curvature" because at the time it was not clear if these nebulae observations were ``local peculiarities" or ``genuine expansion". 

While it is certainly true that the empirical results helped turn the tide against the static universe, it is also true that the results were by no means conclusive at the time. Indeed, \textcite{ORaifeartaigh:2019uft} has argued that it was precisely this lack of conclusive, robust astronomical data that caused Lema\^itre's work to be initially overlooked, contrary to the usual assertion that his work was too obscure to be taken seriously. In this case, we can see evidence of MIA-type reasoning that led Einstein and Eddington to reject the static universe. They clearly extrapolated the prior empirical successes of GR to cosmological phenomena that did not yet have robust empirical results. Furthermore, they recognized that GR did not admit physically reasonable static possibilities. Applying GR on cosmological scales in an internally and externally consistent manner necessarily demanded a dynamically evolving universe. As we have seen, this reasoning actually influenced their interpretation of the limited empirical results available at the time. 

This is slightly different from the Higgs example because the dynamical universe is a robust and unavoidable prediction of GR. It is not a separate module of the theory in the same way that the Higgs was a separable module that was later introduced for reasons of theoretical and empirical consistency. The connection between early empirical confirmations of GR and the theory or model of a dynamically evolving universe that warrants the inference of non-empirical support comes about from the fact that a dynamically evolving universe can be directly derived from GR itself. That is, a dynamically evolving universe naturally follows from a straightforward, consistent application of GR to regimes beyond where it had been tested.

\subsubsection{The Existence of Gravitational Waves}

The history of gravitational waves (GWs) offers another example of MIA-type reasoning. Initially, Einstein doubted that they could exist at all given the non-existence of a gravitational dipole moment. However, the research community wavered back and forth on this question over the decades. Episodes include Einstein eventually deriving three types of GWs, Eddington demonstrating that two of them were pure coordinate artifacts and casting doubt of the existence of the third type, Einstein and Rosen concluding, but not publishing a paper that argued that GWs did not exist, and subsequent discussions between Einstein, Robertson, and Infeld that resulted in the opposite conclusion. Finally work from Pirani and Feymann provided more convincing theoretical justifications for the existence of GWs that inspired concrete interest in searching for them at the famous 1957 Chapel Hill relativity conference \parencite{Cervantes-Cota:2016zjc}.

In their 1972 annual review, \textcite[p.~336]{1972ARA&A..10..335P} noted that it was regrettable that physicists spent decades doubting the existence of GWs, but that it was now understood that GR ``predicts, unequivocally, that gravitational waves must exist; that they must be generated by any nonspherical, dynamically changing system; that they must produce radiation-reaction forces in their source; that those radiation-reaction forces must always extract energy from the source; that the waves must carry off energy at the same rate as they extract it; and that the energy in the waves can be redeposited in matter...". 

Similar to cosmological solutions of GR indicating that the universe need be dynamical, GW were then understood as a robust prediction of GR and physicists had a high degree of confidence in eventually detecting them. The first two decades of such efforts were met with failure, but the discovery of the Hulse-Taylor binary, along with subsequent analysis of the orbital period decay, pointed to strong indirect evidence for the existence of GWs \parencite{1975ApJ...195L..51H, 1979Natur.277..437T}. This paved the way for LIGO and VIRGO, the next generation of laser-interferometry based GW detectors, which culminated in the first direct detection of GWs in 2016 as these instruments detected a binary black hole merger \parencite{LIGOScientific:2016aoc}.

Discovering GWs required truly enormous investments of time, money, and resources, and in this regard, is similar to the Higgs discovery. However, as was the case with the dynamical universe, GWs are not a later supplemental module of the theory, but can be understood as a straightforward consequence of GR because they are directly derivable from the theory. That is, a consistent application of GR's theoretical and programmatic resources necessarily implies the existence of GWs. GR's tremendous empirical merits again legitimized inferring non-empirical support for other areas of the research programme that had not yet been verified, including to phenomena and regimes that had not yet been experimentally probed. The MIA argument justifies the enormous investment (monetary and temporal) and confidence from the community, despite not having any direct (or indirect) empirical evidence for GWs over multiple decades of somewhat frustrating failures.\footnote{Indeed, this has paid off tremendously as we are only in the very beginning of what promises to be a long and fruitful era of gravitational wave astronomy. Gravitational waves have been and will continue to be used to explore the strong gravity regime, alternative theories of gravity, structure formation in the early universe, astronomical processes via multi-messenger signals, etc. See, e.g.\ \textcite{Barausse:2020rsu, Bailes:2021tot, Baker:2017hug, Wolf:2019hun} for some discussions and applications.}

\section{MEA Programme and Cosmological Inflation} \label{s4}

\noindent The theory of cosmological inflation was initially proposed by \textcite{Guth:1980zm, Starobinsky:1980te}. In its modern presentation\footnote{Here I follow standard presentations of cosmic inflation. See e.g.\ \textcite{Baumann:2022mni, Baumann:2009ds, Weinberg:2008zzc, Mukhanov:2005sc} for details.}, the basic idea is that very early in the universe's history, the matter-energy content of the universe was dominated by a scalar field. This scalar field has certain properties (i.e.\ its potential energy is the dominant contribution to the energy density and the potential has a functional form that is relatively flat) that effectively lead to an equation of state that generates a significant repulsive force in the form of negative pressure. This causes a rapid, exponential expansion of the universe's scale factor (`cosmic inflation'!). 

Inflation initially gained traction due to its ability to help resolve perceived explanatory problems in the standard hot big bang model and to provide a causal mechanism for generating density perturbations in the early universe (more on these issues in \S\ref{UEC}). Furthermore, as evidence from successive cosmological probes accumulated, these observations painted a picture of the universe that looks very much like the universe we would expect to see if inflation had occurred. Among other predictions, an inflationary epoch suggests that the universe should have the following properties \parencite{Guth:2004tw}:
\begin{enumerate}
    \item Geometrically flatness
    \item Approximate uniformity in the distribution of its mass-energy content
    \item Possess density perturbations with very particular statistical properties: nearly scale-invariant (i.e.\ approximately independent of length scale), Gaussian (i.e.\ normal distribution), and adiabatic (i.e.\ independent of matter-energy species).
\end{enumerate} 
Briefly, we can understand that these predictions naturally follow from inflation because a period of rapid spatial expansion will dynamically flatten the universe and smooth out inhomogeneities. The properties of density perturbations are a bit more involved, but they follow from treating the scalar field responsible for inflation quantum mechanically.\footnote{See \textcite{Baumann:2009ds} for an excellent pedagogical discussion.} These density perturbations are crucial to our description of the universe because they generate cosmic structure (i.e.\ clusters, galaxies, stars etc). Measurements from WMAP and Planck indicate that the universe we observe in the Cosmic Microwave Background (`CMB') is in very close agreement with these `generic predictions' \parencite{WMAP:2003elm, Planck:2018vyg, Guth:2013sya}.

While this is certainly a nice story so far, it is no surprise that things get a bit more complicated. One complication is that there is a truly enormous variety of inflation models, many with wildly diverging physical motivations, implications for cosmology and particle physics, and empirical predictions. One of the most comprehensive surveys in the literature counts 74 different models of ``simple" single-field inflation \parencite{Martin:2013tda}, \textit{not} including many other more complicated scenarios such as multi-field inflation. In other words, inflation is more of a framework or paradigm than an actual theory. In practice, actually confirming one of these models will prove to be difficult. While the predictions listed above are very generic within the inflationary paradigm, there is significant variety in other predictions that come from different inflationary models. Most importantly, in addition to density perturbations, inflation also produces tensor perturbations in the form of primordial gravitational waves. The crucial observational signature for these tensor perturbations is the so called `tensor-to-scalar ratio' $r$ \parencite[Sect.~3]{Baumann:2009ds}, which measures the ratio of amplitudes between tensor perturbations (i.e.\ primordial gravitational waves) and scalar perturbations (i.e.\ density perturbations of matter). Measuring this quantity (as well as others such as the `spectral index' $n_s$, which quantifies the exact scale dependence of the scalar perturbations) provides an excellent opportunity to discriminate between distinct models and potentially provide further evidence for inflation.\footnote{See Figure 124 in \textcite{Martin:2013tda} for comparisons between many models and their observational predictions, along with arguments that cosmological data is good enough to discriminate amongst them. This can be contrasted with theoretical proposals for dark energy, which also utilize similar kinds of scalar field models to account for the accelerating expansion of the universe in the current epoch. In dark energy model building, there are large sections of the parameter space for the key observables ($w_0$, the value of the dark energy equation of state now, and $w_a$, the time evolution of the dark energy equation of state) where many distinct models are entirely degenerate \parencite{Wolf:2023uno}.}

Unfortunately, despite significant efforts to detect this quantity (which included a prominent false positive \parencite{Cowen2015GravitationalWD}), primordial gravitational waves have not been detected, with Planck placing upper bounds on $r$ ($r < .10$ \textcite{Planck:2018vyg}), disfavoring many of the simplest inflationary models. While there are many surviving models and the paradigm is still considered to be on very strong footing by the majority of the physics community \parencite{Chowdhury:2019otk, Guth:2013sya}, the class of models favored following these results are known as ``plateau inflation" \parencite{Martin:2013nzq}. Such models have been criticized by skeptics of inflation as being somewhat less appealing considering that they require more parameters to generate the relevant observables and require more finely-tuned initial conditions to get the inflation off the ground \parencite{Ijjas:2013vea}. 

While not yet a full-blown crisis, the somewhat unexpected, persistent difficulty in nailing down direct empirical support for a particular inflation model and the trend towards seemingly more complicated models based on the constraints we do have, has caused a minority of physicists to question the status of the inflationary paradigm. Furthermore, it seems like truly \textit{conclusive} empirical tests of inflation are currently beyond our experimental capabilities, which has led Dawid to the conclusion that ``assessments of the status of inflationary cosmology will most probably have to rely heavily on assessments of scientific underdetermination for a long period of time" \parencite[p.~91]{Dawid2013-DAWSTA}. In a more recent article that attempts to adjudicate some of the recent debates between members of the physics community who favor inflation and those who have wavered, \textcite{DawidManuscript-DAWIIC} more explicitly argue that we should turn to MEA to assess inflation given the current status of the empirical picture. Following this suggestion, I will engage with the MEA assessments offered by Dawid and McCoy as well as provide my own MEA assessment for inflation, beginning first with NAA and UEA, before proceeding to MIA.

\subsection{No Alternatives Argument} Inflation is the dominant paradigm in early universe cosmology and is generally considered to be the best theory for providing robust, convincing explanations for the various problems encountered earlier. Yet, recent difficulties with the paradigm have caused some notable members of the community to approach alternatives with renewed energy. The most prominent alternatives in the literature at present are bouncing cosmologies, with a particular emphasis on ``ekpyrotic" bouncing cosmologies \parencite{Steinhardt:2001st, Steinhardt:2002ih, Ijjas:2018qbo, Ijjas:2019pyf}. Other bouncing alternatives include the Matter Bounce, String Gas cosmology, and the Pre-Big-Bang scenario \parencite{Brandenberger:2016vhg}. 
    
All of these alternatives \textit{can} solve the problems that inflation addresses, the question rather becomes at what cost do these models solve these problems and do they represent reasonably adequate, viable alternatives? Here we appeal to \textcite{Kuhn1977-KUHOVJ} and his model of theory choice, whereby scientists weigh objective theory virtues such as empirical accuracy, scope, and simplicity/parsimony according to their own subjective preferences. 
    
Briefly, consider how an ekpyrotic model handles issues we have seen like homogeneity, flatness, and scalar density perturbations. An ekpyrotic model induces contraction within the universe through a scalar field with a steep, negative exponential potential (rather than a flat, positive potential). It turns out that slowly contracting universes lead to remarkably similar outcomes as those expected from rapidly expanding universes, as the slow contraction dynamically flattens the universe and produces uniformity in the matter-energy distribution.\footnote{See \textcite{Ijjas:2018qbo} for explanation on contraction dynamics and a comparison with inflationary dynamics.} However, when contraction reverses and an expansion phase similar to the one we are currently experiencing, these models often encounter catastrophic ``ghost instabilities". These particularly nasty pathologies result from an unbounded Hamiltonian that both renders the theory perturbatively ill-defined and allows for the infinite production of particle states at arbitrarily high energies \parencite{Rubakov:2014jja, Wolf:2019hzy}. As cosmological models with such pathologies are not considered to be phenomenologically viable, proponents of bouncing models have found ways to avoid them by introducing more complicated dynamics by modifying gravity \parencite{Cai:2017dyi, Ijjas:2016vtq, Easson:2011zy}. Additionally, ekpyrotic models can produce the observed, nearly scale-invariant spectrum of scalar density perturbations, but it turns out that an additional scalar field is needed to realize this \parencite{Brandenberger:2016vhg}. Proponents of inflation have argued that these are serious problems, and if they can be overcome at all, require inordinately difficult and poorly motivated modifications to evade them \parencite{Linde:2014nna, Kallosh:2007ad, Linde:2009mc}. \textcite{WolfThebault} have identified this kind of ``dynamical fine-tuning" as a major reason why physicists' tend to have a general distaste for such bouncing models.
    
Despite these complications that call into question the parsimony of such models, bouncing models do have some advantages over their inflation competitors. For example, \textcite{Penrose:1988mg, Hollands:2002yb} have forcefully argued that the inflationary paradigm encounters a significant conceptual problem. Any universe emerging out of an initial singularity would naturally be expected to possess a very high entropy. Furthermore, a state with entropy this high would be incompatible with the occurrence of inflation while the possible initial states that could be compatible with inflation seem to be vanishingly small in comparison. On the contrary, these ekpyrotic models are constructed to be non-singular and thus evade this specific issue concerning the entropy. Another advantage comes about from the fact that these kinds of bouncing models predict that there should not be significant production of primordial gravitational waves \parencite{Ijjas:2018qbo}, meaning that they are more easily made consistent with the aforementioned Planck results. 
    
The significance of all these issues for both paradigms is an open issue and actively debated in the physics community. Furthermore, the philosophy literature has waded into this debate as well, with \textcite{DawidManuscript-DAWIIC} providing a philosophical analysis of the ongoing debate and \textcite{WolfThebault} offering a philosophical analysis comparing the explanatory merits of both approaches. Ultimately though, resolution will come from the technical results and subjective judgment of individuals within the theoretical physics community. In other words, the ultimate outcome of the NAA for inflation ``must play out largely among the physicists involved in the corresponding research programs" \parencite[p.~28]{DawidManuscript-DAWIIC}. The takeaway though, is that at present there are actively pursued alternatives that are empirically adequate, viable options, even if the judgment of the community as a whole still (understandably) finds inflation to be more desirable according to their theory choice preferences. With the current status quo, it does not seem like one can draw any firm conclusions regarding the status of an NAA argument in the context of early universe cosmology.

\subsection{Unexpected Explanatory Coherence Argument}\label{UEC} On the other hand, the UEA argument for inflation is particularly strong. Indeed, all of the generic predictions listed in the beginning of \S\ref{s4} can be plausibly understood as contributing to this argument. The original proposals for inflation emerged from exploring the consequences of ideas in high energy particle physics and applying them to a cosmological context. For example, \textcite{Guth:1980zm} was investigating the cosmological consequences of phase transitions in grand unified theories, while \textcite{Starobinsky:1980te} was investigating potential contributions from quantum mechanical corrections to GR in the form of higher order curvature terms that could be relevant in the high energy density environments of the early universe. They found that their investigations pointed to the conclusion that the universe underwent an inflationary period of exponential (`quasi-de Sitter') expansion.

However, it was Guth who quickly realized that the dynamics resulting from the phase transitions he was investigating could offer a resolution to the classic puzzles of standard hot big bang cosmology. Among these, the \textit{horizon problem} and the \textit{flatness problem} stand out.\footnote{Again, following standard presentations of the subject found in \textcite{Baumann:2009ds, Baumann:2022mni, Weinberg:2008zzc, Mukhanov:2005sc}.} The \textit{horizon problem} refers to the fact that the universe is remarkably uniform over large scales, to the point that CMB measurements have revealed that the universe has a uniform temperature of 2.73 K, with average variations of 1 part in 100,000 across the sky. The scales that homogeneity holds over are so large that the vast majority of the universe is not within a common causal horizon, leaving the question of how distant points, points that do not share a causal past, could display such remarkable uniformity. The \textit{flatness problem} refers to measurements that indicate that the universe is nearly spatially flat today. This is surprising because an exactly flat universe corresponds to a very particular critical density value for the matter-energy content in the early universe, with any deviation from this value being an unstable fixed point that would lead the universe to rapidly diverge from spatial flatness. The fact that the universe is still so remarkably close to flatness today indicates that this critical density value needed to be extraordinarily special.

Inflation, despite not being specifically designed to solve these problems, immediately provides compelling explanations for these observations. Inflation resolves the horizon problem and explains large scale uniformity. It indicates that the universe actually did have a common causal past; such extreme expansion makes it only \textit{appear} today as if very distant points are outside each other's past light cones. Furthermore, exponential expansion smooths out any inhomogeneities and produces uniformity. The flatness of the universe is also no longer a mystery because this extreme stretching of space will naturally flatten the universe, almost regardless of what its initial curvature or matter-energy density was.

Perhaps inflation's most important accomplishment, if its turns out to be correct, is providing a theory that explains the origin of density perturbations. Before inflation, cosmologists could provide a purely phenomenological account of cosmic structure, but there was no predictive theory of its origin or properties \parencite{Smeenk:2018dbt}. However, it was also quickly realized that inflation provides a compelling origin story for these density perturbations in the form of a causal mechanism that generates them (i.e.\ tiny quantum mechanical variations in the inflation field) \parencite{Mukhanov:1981xt, Hawking:1982cz, guthpi1982, bardeen1983}. Inflation remarkably connects the observed large-scale structure of the universe to tiny quantum fluctuations during this period of inflation. As the theory was not engineered to produce this result, inflation unexpectedly provided a powerful, causal, and remarkably coherent explanation of the observed large-scale structure of the universe. ``Rather than pulling the initial spectrum out of a hat, as one might suspect of the earlier proposals, the inflationary theorist can pull a [nearly scale-invariant] spectrum [...] out of the vacuum fluctuations of a quantum field" \parencite[p.~9]{Smeenk:2018dbt}. 

Indeed, alongside the prediction of spatial flatness, this incredible result not only counts as an instance of UEA, but can also be understood to be a powerful instance of predictive novelty. Inflation theorists were able to use the theory to predict important features of universe such as its spatial flatness and the properties of primordial fluctuations (e.g.\ their deviations from scale-invariance and statistical properties) long before they were actually observed \parencite[Sect.~7]{WolfDuerr}.
    
Another potential instance of UEA, pointed out by \textcite{DawidManuscript-DAWIIC}, concerns the potential for inflation to provide an explanation for the value of the cosmological constant. As the inflation paradigm strongly implies a multiverse framework known as `eternal inflation' (see \textcite{Guth:2007ng, Aguirre:2007gy} for a physics review of eternal inflation and its implications), it can be argued that this framework, in conjunction with anthropic reasoning, provides the most reasonable explanation for the particular value of the cosmological constant. While the merits of this instance of UEA will no doubt depend on one's thoughts regarding the multiverse issue and the validity of anthropic reasoning, this serves as another potential example of UEA for the inflationary paradigm.

\subsection{MIA Argument and Cosmological Inflation} \label{s5}

As mentioned before, MIA is an empirical argument that relies upon connecting the hypothesis or theory that one would like to infer meta-level support for to the actual empirical evidence that one is citing in this inference. And in doing so, we need ``a reasonable understanding as to what can count as a prediction of a similar kind". The theory of cosmological inflation is a natural outgrowth of both particle physics and general relativity, so it makes sense to consider if the inference made in support of inflation is of a similar kind to other successful inferences within these research programmes. For the Higgs mechanism, this inference between empirical support and theory relied upon the role that the Higgs mechanism played resolving an important inconsistency between observations and the standard model in a manner that was integral to achieving empirical adequacy. For the examples from cosmology, the inferences relied upon consistently applying GR (an empirically successful theory) to regimes and phenomena that had yet to be tested.

At first glance, the introduction of inflation into the standard model of cosmology seems most similar to the introduction of the Higgs mechanism into the standard model of particle physics. The inflaton field and the Higgs field are both scalar fields built using standard quantum field theory machinery, and they were both attached to their respective theories in order to solve outstanding problems and provide explanations for observations that were puzzling given the theoretical frameworks the respective communities were operating within. Furthermore, there is not a sense in which a field that serves the specific role or has the particular properties that we expect of the inflaton is directly derivable from established knowledge in either the standard model of particle physics or the standard model of cosmology. For example, trusting in the merits of GR following significant empirical confirmations necessarily entails that the universe evolves dynamically and that GWs exist because these are straightforward consequences of the framework. Inflation does not follow this pattern in any sense when viewed from either the particle physics or cosmology angle. Like the Higgs mechanism, the inflation hypothesis is an addition to the standard model of cosmology. Thus, it is most natural to compare the MIA argument for inflation to the MIA argument for the Higgs mechanism.

\textcite{DawidManuscript-DAWIIC} briefly sketch an MIA argument for cosmological inflation by arguing that MIA support for inflation can be identified in both of the particle physics and cosmology research traditions. On the particle physics side, they argue that the predictive success of particle physics and its underlying principles instills trust in our abilities to construct successful scalar potentials within the framework of quantum field theory. This presumably refers to the historically successful example of the Higgs field (an empirically verified scalar field) and to the fact that fully nailing down any serious candidate for the inflaton will involve probing this field's interactions with established particle physics knowledge. From cosmology, they point to Peebles' identification of previous instances of successful non-empirical assessment in cosmology, instances that involved trusting GR on far larger distance scales than had ever been empirically probed before \parencite{Peebles+2020}. They argue that this licenses inflation theorists to trust the conjunction of GR and particle physics inspired inflatons at the energy scales relevant to inflation and the very early universe. 

Unfortunately for inflation, these general arguments do not quite work because there is an important disanalogy between a theoretical extension like inflation and a theoretical extension like the Higgs mechanism. As we have seen, the Higgs mechanism is not merely an important component of the standard model of particle physics, but rather it is \textit{essential} for the empirical adequacy and consistency of the whole model. Can the same be said about inflation and its relationship to the standard model of cosmology? In other words, do the demands of both empirical adequacy and consistency \textit{necessitate} the introduction of inflation to the standard model of cosmology?

The answer, which may be surprising to some given the amount of attention that inflation receives, is most certainly not, and this is readily acknowledged by both physicists and philosophers. As \textcite[p.~25]{Baumann:2009ds} notes, ``the flatness and horizon problems are \textit{not} strict inconsistencies in the standard cosmological model" and that with the right initial conditions for the density parameter and inhomogeneities the big bang model certainly accounts for the present observational picture (see also \textcite[p.~184]{guth1997inflationary}). \textcite[p.~13]{Earman1999-EARACL}, in an early philosophical assessment of inflation, likewise conclude that ``the basic motivation for the inflationary paradigm comes from alleged inadequacies with the standard big bang model. It is far from clear to us that these are indeed genuine difficulties since they have to do not with empirical adequacy but with styles of explanation".

Earman and Mosterin point out that the explanations afforded by the big bang model are not obviously deficient according to the standard accounts of explanation. Additionally, they argue that these explanations are perfectly coherent with the most influential account of explanation in the literature: the deductive-nomological (D-N) model \parencite{Hempel1948-HEMSIT-5}. This model holds that for a given explanandum, a sufficient explanans results from the combination of appropriate initial conditions and dynamical laws. We can use both the currently observed conditions in the universe and the dynamical Einstein field equations to retrace the evolution of the universe and determine the unique initial conditions immediately after $t_i=0$. We can then demonstrate how these initial conditions produce the universe we see today by rolling them forward to $t_f$ again by applying the dynamical Einstein equations. This is a perfectly acceptable explanation in virtually any other dynamical problem in the physical sciences. Why should standard applications of initial conditions and dynamical laws be so egregiously problematic in cosmology?

Before proceeding, it is important to emphasize that just because a theory or model is empirically adequate or can be understood to satisfy some philosophical model of explanation, this does not mean that it is not deficient in some way. Physicists are clearly dissatisfied with the big bang model because of the \textit{fine-tuning} of initial conditions needed to make it work. Indeed, fine-tuning can amount to an empirical problem that rises to the level of an inconsistency that demands a resolution. However, there are many different types of fine-tuning and it is important to distinguish examples of fine-tuning that simply are contingent facts about the way things are as opposed to those that demand some kind of important explanatory resolution. As \textcite{Hossenfelder2019-HOSSFE} has emphasized, there are many assumptions within our theories, ranging from the particular values certain dimensionless parameters take (among all possible values) to the particularly sets of mathematically consistent axioms we use (among an infinite number of choices), that are there simply because they adequately describe nature. No one denies that it is appealing to find explanations for these sorts of things and that it is worth searching for them, but they do not necessarily ``scream for explanation" in the same manner that the most egregious instances of fine-tuning do.    

When exactly does fine-tuning ``scream for explanation"? \textcite{Hossenfelder2019-HOSSFE} argues that fine-tuning screams for an explanation precisely when we have a well-defined probability distribution that allows us to clearly quantify the unlikeliness of what we are observing. For example, if we observed a significant deviation from the predictions of thermodynamics, a standard D-N explanation that the combination of initial conditions and dynamics just happened to produce an unlikely state would obviously be deficient. Observing gross violations of the Born rule would be similarly shocking. These instances would demand some further explanation because this kind of fine-tuning becomes an \textit{empirical problem} that can be cashed out in terms of careful probabilistic reasoning that empirically predicts what we should (and should not) be observing. 

Does the big bang model exhibit this kind of fine-tuning? \textcite{McCoy2015-MCCDIS, McCoy2018-MCCTII-2} has explored this question in some detail, and argued that the physics literature largely lacks substantive justification for its interpretation of these initial conditions as ``improbable". Additionally, arguments that these conditions are truly unlikely or egregiously problematic must rely on defining a suitable probability measure. Such probability measures in cosmology cannot be defined without introducing arbitrary cut-offs to regularize for divergent integrals and furthermore, the resulting verdict is very sensitive to such arbitrary choices \parencite{Schiffrin:2012zf}. These problems are intimately related to the infinite dimensional phase space of GR (even in the case of the ``minisuperspace" that, roughly speaking, restricts itself to FLRW spacetimes). It is not clear that a probability measure can be chosen in any physically meaningful way \parencite{Curiel:2015oea}.

Given that this clear line of probabilistic reasoning is blocked, one can appeal to more qualitative analyses of fine-tuning. For example, \textcite{baras-ms} argues that fine-tuning calls for an explanation when it instantiates an \textit{extraordinary} type, which can be understood as a fact that is contrary to what we would expect given our background knowledge. One can easily argue that this characterization applies to the problems in the big bang model given its issues with dynamical instability and causality, as these immediately strike us as deeply puzzling, if not unsettling. However, this certainly does not rise to the level of an inconsistency that impinges upon the empirical adequacy of the model as we saw in the example of the Higgs mechanism. This does not deny that inflation offers significant gains in coherence and explanatory power (see e.g.\ \textcite[Sect.~4]{WolfDuerr}; \textcite[Sect.~3\&4]{WolfThebault})\footnote{Coherence and explanatory power also play a significant role in other debates within contemporary cosmology. See \textcite{duerr-wolf-forthcoming} for an analysis of the MOND/Dark Matter debate with a particular focus on the coherence (and ad-hocness) of these proposals.}; however, such merits (as tremendous as they are) are arguably more plausibly viewed as being indicators of reasons to \textit{pursue} a theory \parencite{Laudan1977-LAUPAI, Seselja2014-EELEJI, Nyrup2015-NYRHER}, rather than as indicators of \textit{confirmation}.

Despite inflation being a natural extension of both standard models of particle physics and cosmology, making an MIA-style argument for inflation would require inferring support for inflation primarily on the basis of our explanatory preferences, rather than on the empirical success of and consistency with well-established theories that would be cited in such support. This does not mean that MIA arguments cannot work in instances of fine-tuning, but rather that it is important to consider whether the fine-tuning present represents something that can be reasonably characterized as a contingent fact or represents some kind of empirical failure or inconsistency in the current standard model. 

The empirically observed state of the universe does not warrant a significant MIA inference in favor of inflation as it did in the Higgs example precisely because this observed state cannot be understood to represent an \textit{inconsistency} with the current standard model that bears upon the empirical adequacy of the framework; without such an inconsistency, an MIA-type inference to inflation lacks the bite that the inference to the Higgs mechanism so clearly has. Furthermore, the empirical and theoretical evidence one needs to cite in such an inference simply does not \textit{demand} a resolution with the same urgency or point to a unique solution with the same force. Nor can inflation be understood to be a directly derivable consequence from an empirically well-established theory as we saw in the examples of the dynamically evolving universe and the existence of gravitational waves. In these previous examples of MIA, consistency arguments directly tied to the empirical evidence itself necessarily implicated the unconfirmed theories we inferred MIA support for. The same cannot be said for inflation. Inflation's explanations are indeed far better and more satisfying, but the aesthetic appeal of inflation's explanatory merits is more properly understood as a (very compelling!) reason to further pursue the theory \parencite{WolfDuerr}.\footnote{\textcite{Cabrera2021-CABSTN} has argued that MEA should be reworked as a programme that belongs to the context of pursuit rather than justification or confirmation. I am sympathetic to this argument. As indicated in this section, I think that the MEA case for inflation would be significantly stronger if the MEA arguments were construed as arguments for pursuit.}

Applying MIA to inflation is not very convincing despite offering significant explanatory advantages over the standard model without inflation. Along with an inconclusive NAA argument, there is not a convincing case for ascribing significant non-empirical confirmation to the theory of cosmological inflation because two of the three arguments needed to generate significant limitations to scientific underdetermination are fairly weak. 

%This does not mean that the MIA argument, and MEA as whole, cannot eventually be used as a methodological tool to support inflation. In the following section, I will explore some scenarios that would make an MIA argument, and by extension a full MEA case, far more compelling for inflation.

\section{Future Meta-Empirical Prospects for Inflation} \label{s6}

\noindent To this point, we have seen that successful examples of MIA-type reasoning for a particular hypothesis in cosmological and particle physics contexts have followed two patterns. (i) The hypothesis is inextricably tethered to both the empirical adequacy and consistency of the larger research programme. (ii) The hypothesis is a directly derivable consequence of an underlying theory or framework that has robust empirical support. As I have argued, inflation is a poor fit for (i) because inflation is not required for the consistency or empirical adequacy of the rest of the cosmological research programme, in marked contrast to the role that the Higgs mechanism plays in the particle physics research programme. Additionally, (ii) does not work either because neither of the standard models can be understood to necessarily entail inflation, as inflation is rather a separable extension of these programmes. 

However, this does not mean that future developments do not have the potential to make an MIA-style argument for cosmological inflation more compelling. Indeed, there are conceivable future scenarios that could bring inflation into closer alignment with either (i) or (ii), and in doing so render inflation ``a prediction of a similar kind", where the empirical considerations more directly warrant an inference for an inflationary epoch. For example, there is arguably a scenario where further empirical observations would make the relationship that inflation has to the standard model of cosmology more analogous to the relationship between the Higgs and the standard model of particle physics. This involves the aforementioned tensor-scalar ratio $r$ and a positive detection of primordial gravitational waves. 

Detecting these primordial gravitational waves, or tensor perturbations, is considered to be a holy grail for cosmology because measuring their properties has the potential to put powerful constraints on early universe theories. One reason for this is that such gravitational waves would be measured through detecting a so-called `B-mode' pattern of polarization, which is significant because it can be proven that \textit{only} tensor perturbations produce B-mode polarization, whereas scalar perturbations \textit{only} produce E-mode polarization \parencite{Kamionkowski, Zald}. The upshot is that there are only a few known processes that could produce such signals, and the processes that could produce these signals are clearly delineated from those that create the scalar perturbations we have already measured. 

While a positive detection would clearly point towards one of these processes, the mere detection of primordial gravitational waves would not uniquely single out inflation as is commonly believed. Indeed, \textcite{Brandenbergerbmodes} notes that such signals can be produced both in some versions of bouncing cosmologies (depending on details of the contraction phase before the bounce) and even in standard big bang cosmology (through global phase transitions known as cosmic strings). However, the particular properties of these gravitational waves very well could eliminate these candidates. For example, bouncing models that produce significant amounts of gravitational waves also would induce particular non-Gausianities in the statistics of the signal \parencite{Brandenbergerbmodes}, while gravitational waves sourced by cosmic strings in standard big bang cosmology would vanish over superhorizon scales due to causality constraints \parencite{Baumann:2009mq}. Thus, a positive detection of primordial gravitational waves, combined with the right properties in the gravitational wave spectrum, could very well point to an inflationary epoch and even indicate the energy scale of such a period, as well as constrain the shape of the inflaton potential \parencite{Baumann:2009ds}. 

In this scenario, there would be a much stronger case that the standard model of cosmology \textit{must} invoke cosmological inflation to maintain its empirical adequacy (see also \textcite{Smeenk:2017uof} for philosophical analysis along similar lines). After all, the presence of empirical signatures that cannot be produced in any known way in standard big bang cosmology would seemingly render the standard model empirically inadequate in its account of the observations, and indicate that an extension such as inflation is necessary. Such a scenario would arguably resemble the kinds of consistency arguments we have already seen much more closely. Furthermore, this would also likely eliminate alternative theories and create a much stronger NAA argument as well. 

Some might even be tempted to say that this would constitute direct proof for the inflaton. Yet, the inflaton is most often viewed as an additional matter field that falls within the purview of standard quantum field theory and particle physics, and this should inform our standards of direct proof. Traditionally, most other fields in particle physics have been probed with deep inelastic scattering experiments, whereby they are directly `observed' via particle traces extracted from the detectors and inferred through the decay products resulting from such collisions. However, as is detailed by \textcite{Dawid2015-DAWHDA}, the actual process of confirming the Higgs boson involved many interesting complications due to the fact that it is electrically neutral and consequently does not produce a trace in the detector. This meant that subsequent decay events could not be uniquely attributed to a single vertex in a scattering event as observations allowed for multiple possible explanations for the observed decay products. This required the existence of the Higgs boson to be established statistically based on examining the number of events observed vs. what would be expected in a background without the Higgs. Confirming the inflaton via traditional particle physics experimentation would not merely be far more challenging than these complications with the Higgs, but is actually considered by most to simply not be within the realm of possibility due to the $\sim 10^{15}$ GeV energies inflation is believed to have occurred at; energy scales that lie far outside the realm of terrestrial collider experiments. 

If $r$ could be measured and the inflaton potential and energy scale could be meaningfully constrained, this may force a re-assessment of what it means for a particle or field to be confirmed experimentally. However, even if it is not realistic to confirm the inflaton in the \textit{exact} same manner as we did the Higgs or other standard model particles, there are proposals within the field of cosmology that would offer something fairly analogous. 

For example, exploring the physics of reheating would be a significant step towards this goal. Reheating refers to the epoch immediately after inflation, where the inflaton field oscillates around the minimum of its potential, decays and transfers its energy into the creation of other particles, and thereby produces the matter-energy content that populates the universe during the early radiation dominated stage. The details of this physics are highly sensitive to the interaction between the inflaton and other standard model fields \parencite{Martin:2014nya}. Thus, we would gain invaluable information regarding how the standard model fields we know couple to and interact with the inflaton, and how these interactions and byproducts of reheating inform our expectations for early universe observables. Just as exciting, a somewhat new avenue in theoretical cosmology known as ``cosmological collider physics" has recently emerged, with the idea being that the inflationary epoch might have excited entirely new fields and created particles we are not familiar with that have masses comparable to the Hubble scale \parencite{Arkani-Hamed:2015bza}. The idea is that these interactions and subsequent decays would leave statistical imprints on the scalar perturbations in the CMB, in analogy with how collider experiments leave statistical imprints and patterns that can be measured on detectors. While the details of all of these proposals have yet to be fully ironed out, it is not inconceivable to think that they could offer direct empirical confirmation of inflation in a manner somewhat analogous to the way in which other fields and particles are understood to have received direct empirical verification. 

Before such direct proof could be obtained though, inflation could plausibly be understood to have a much stronger argument for non-empirical confirmation under the scenario explored at the beginning of this section, where the detection of primordial gravitational waves indicates, as a matter of necessity, that such an inflationary epoch must be grafted onto the standard $\Lambda$CDM model to resolve an important empirical inconsistency.

\section{Conclusion} \label{s7}
\noindent Meta-empirical assessment and confirmation is an interesting new methodological tool in the philosophy of science. There is evidence that this type of reasoning has played a significant role in developing the standard models of both particle physics and cosmology; however, meta-empirical considerations will become even more important in the future as theorizing becomes ever more detached from timely empirical exploration. The theory of cosmological inflation is a fascinating addition to the standard model of cosmology that has the potential to unify phenomena at large and small scales, as well as provide compelling explanations for the observed state of the universe. However, as of now, inflation does have viable alternatives, limiting the appeal of an NAA-type argument. Additionally, inflation fails to support a compelling MIA-type argument because the current state of empirical observations does not implicate inflation to the same degree seen in other relevant successful instances of MIA. Future observations could very well pave the way for a more compelling non-empirical confirmation case for inflation. For the time being though, lacking two of the three main pillars of the MEA programme, it is premature to ascribe a strong degree of non-empirical confirmation to inflation. 

\section*{Acknowledgements} 
\noindent I am grateful to Richard Dawid, Karim Th\'ebault, James Read, Patrick Duerr, and Michael Townsen Hicks for many insightful discussions and detailed feedback on this project. I also am grateful for support from St. Cross College, University of Oxford.

% \bibliographystyle{dcu}
% \bibliography{meta}
\printbibliography

\end{document}